\documentstyle[preprint,mbf,wrapft]{ptptex}

\title{
Higgs Field as Weak Boson in five Dimensions
}

\author{
Kohzo {\sc Nishida}
\footnote{E-mail: EZF01671@nifty.ne.jp} 
}

\inst{
Department of Physics, Kyoto Sangyo University, Kyoto 603-8555, Japan
}
\abst{
We propose a five-dimensional standard model which regards the Higgs field as 
a weak boson associated with the fifth dimension. The kinetic term of the 
Higgs field is obtained from the fifth components of field strengths defined 
in five dimension. The coupling constant of the fermion fields and the Higgs 
field is only the weak coupling constant. However, since the vacuum expectation 
value depends on the fifth coordinate, we can explain the various mass spectrum 
of elementary particles.
}

\begin{document}

\maketitle

Some models that regard the Higgs field as the gauge field have been studied.  
Hatanaka, Inami and Lim\cite{rf:1} have proposed higher-dimensional gauge 
theories to solve the gauge hierarchy problem by invoking higher-dimensional 
gauge symmetry and they have used an idea that regards the extra space 
component of the gauge field as the Higgs fields in it. 

The purpose of this paper is to construct a five-dimensional standard model 
based on their idea. We identify the higgs particle with a weak boson associated 
with the fifth dimension. The kinetic term of the Higgs field is given by 
$F^{\mu 4}F_{\mu 4}$ obtained from five-dimensional field strengths 
$F^{\alpha \beta}$$(\alpha,\beta=0,1,2,3,4)$. The coupling constant of the fermion 
fields and higgs field is only the weak coupling constant $g_2$. However, since 
the vacuum expectation value depends on $x_4$, we can explain the various mass 
spectrum of elementary particles.

Now let us consider Dirac $\gamma$-matrices for five-dimensional Lorentz group.
We introduce
\begin{equation}
               \gamma^{4} \equiv
               i\gamma^{5}=i(i\gamma^{0}\gamma^{1}\gamma^{2}\gamma^{3}) 
\end{equation}
for the fifth dimension and define
\begin{equation}
               \gamma^{\alpha}=(\gamma^{\mu},\gamma^{4}),\quad \alpha=0,1,2,3,4,
\end{equation}
where Greek letters such as $\alpha$ and $\beta$ label five-dimensional 
space-time vectors, Greek letters such as $\mu$ and $\nu$ label 
four-dimensional space-time vectors. Clearly, they follow that 
\begin{equation}
\label{eqn:a}
               \{ \gamma^{\alpha},\gamma^{\beta} \} = 2g^{\alpha \beta}, \quad
               g^{\alpha \beta} = {\rm diag}(+1,-1,-1,-1,-1), 
\end{equation}
\begin{equation}
               \gamma^{\alpha\dagger} = \gamma^{0}\gamma^{\alpha}\gamma^{0}.
\end{equation}
For a proper infinitesimal Lorentz transformation 
$\Lambda^{\alpha}_{\ \beta}=\delta^{\alpha}_{\ \beta}+\varepsilon^{\alpha}_{\ \beta} $,
an $S$ which satisfies 
$S^{-1}\gamma^{\alpha}S=\Lambda^{\alpha}_{\ \beta} \gamma^{\beta}$ is
\begin{equation}
               S=1-\frac{i}{4}\sigma_{\alpha\beta}\varepsilon^{\alpha\beta},
\end{equation}
where
$\sigma^{\alpha\beta}=(i/2)(\gamma^{\alpha}\gamma^{\beta}-\gamma^{\beta}\gamma^{\alpha} ) $.
We can show that $S^{-1}=\gamma^{0}S^{\dagger}\gamma^{0}$.

By using $\gamma^{\alpha}$, we have the fermionic part of the Lagrangian density:
\begin{equation}
               {\cal L}_{f} = i \bar{\Psi}\gamma^{\alpha}D_{\alpha} \Psi.  
\label{eqn:fermion}
\end{equation}
In chiral spinor notation, this is:
\begin{eqnarray}
               {\cal L}_{f}
               & =& i 
               \left(
               \begin{array}{cc}
                              \psi^\dagger_L,  & \psi^\dagger_R 
               \end{array}
               \right)
               \left(
               \begin{array}{cc}
                             \bar{\sigma}^\mu D_{\mu} & iD_{4} \\
                              -iD_{4}       & \sigma^\mu D_{\mu}
               \end{array}
               \right)
               \left(
               \begin{array}{c}
                              \psi_L\\
                              \psi_R 
               \end{array}
               \right)    \nonumber \\
               & =& i \bar{\Psi} \gamma^{\mu} \left(
               \begin{array}{cc}
                              D_{\mu} & 0 \\
                              0          & D_{\mu}
               \end{array}
               \right)\Psi
               +i \bar{\Psi} \gamma^{4} \left(
               \begin{array}{cc}
                              0      & D_{4} \\
                              D_{4} &  0
               \end{array}
               \right)\Psi,
\end{eqnarray}
where $\sigma_\mu = (1,\mbox{\boldmath $\sigma$})$,  
$\bar{\sigma}_\mu = (1,-\mbox{\boldmath $\sigma$}),$ and
\begin{equation}
               \psi=\left(
               \begin{array}{c}
                              u \\
                              d
               \end{array}
               \right) \ \mbox{for quarks},
               \quad
               \psi=\left(
               \begin{array}{c}
                              \nu \\
                              e
               \end{array}
               \right) \ \mbox{for leptons.}
\end{equation}

Next, let us demand local $SU(3) \times SU(2)_{L} \times U(1)$ gauge symmetry. 
We denote the Lorentz four-vectors of the gauge fields as 
$\displaystyle A^{(3)}_\mu=\frac{1}{2}\lambda_a\cdot A^{(3)a}_\mu$,
$\displaystyle A^{(2)}_\mu=\frac{1}{2}\tau_a\cdot A^{(2)a}_\mu$,
$\displaystyle A^{(1)}_\mu$, where $\lambda_a \ (a=1,\cdots,8)$ and $\tau_a \ (a=1,2,3)$ 
are the Gell-Mann matrices and the Pauli matrices, respectively.
We introduce the fifth component of the $SU(2)_L$ gauge field $H$:
\begin{equation}
               H=\frac{1}{2}(\tilde{ \phi}, \phi)=
               \frac{1}{2}
               \left(
               \begin{array}{cc}
                              \phi^{0*} & \phi^{+} \\
                              -\phi^{-} & \phi^{0}
               \end{array}
               \right),
\end{equation}
where $\phi$ is an $SU(2)_{L}$ doublet and $\tilde{ \phi}=i\tau_{2}\phi^{*}$ 
is the conjugate of $\phi$. We define covariant derivatives acting on $\Psi$
\begin{eqnarray}
               {\cal D}^{(\varphi)}_{\mu} &=&
               \left(
               \begin{array}{cc}
                              D_{\mu}^{(\varphi)L} & 0 \\
                              0  & D_{\mu}^{(\varphi)R}
               \end{array}
               \right), 
               \label{eqn:Dmu} \\
               {\cal D}^{(\varphi)}_{4} &=& 
               \left( 
               \begin{array}{cc}
                              0 & \partial_{4}-i g_2 H \\
                              \partial_{4}-i g_2 H^{\dagger} & 0
               \end{array}
               \right),
\label{eqn:H}
\end{eqnarray}
with $\varphi=q,l,$ where
\begin{eqnarray}
               D_{\mu}^{(q)L} &=& \partial_{\mu}-ig_2 A^{(2)}_\mu
                                                            -ig_1 \frac{1}{2} Y_{L}^{(q)} A^{(1)}_{\mu}
                                                            -ig_3 A^{(3)}_\mu, \nonumber \\
               D_{\mu}^{(q)R} &=& \partial_{\mu}-ig_1 \frac{1}{2} Y_{R}^{(q)} A^{(1)}_{\mu}
                                                            -ig_3 A^{(3)}_\mu, \nonumber \\
               D_{\mu}^{(l)L} &=& \partial_{\mu}-ig_2A^{(2)}_\mu
                                                            -ig_1 \frac{1}{2} Y_{L}^{(l)} A_{\mu}, \nonumber \\
               D_{\mu}^{(l)R} &=& \partial_{\mu}-ig^{(1)}\frac{1}{2} Y_{R}^{(l)} A_{\mu},
\end{eqnarray}
and
\begin{eqnarray}
               Y_{L}^{(q)} &=& \left(
               \begin{array}{cc}
                              \frac{1}{3} & 0 \\
                              0    & \frac{1}{3}
               \end{array}
               \right),
               \quad
               Y_{R}^{(q)}=\left(
               \begin{array}{cc}
                              \frac{4}{3} & 0 \\
                              0    & -\frac{2}{3}
               \end{array}
               \right), \nonumber \\
               Y_{L}^{(l)} &=& \left(
               \begin{array}{cc}
                              -1 & 0 \\
                              0  & -1
               \end{array}
               \right),
               \quad
               Y_{R}^{(l)}=\left(
               \begin{array}{cc}
                              0 & 0 \\
                              0 & -2
               \end{array}
               \right),
\end{eqnarray}
where $g_k$ are the coupling constants of the gauge fields $A^{(k)}_\mu (k=1,2,3)$, 
and the $q$ and $l$ superscripts indicate that the covariant derivatives act on 
quark and lepton fields, respectively. The fifth components of the $SU(3)$ and 
$U(1)$ gauge fields are forbidden  by gauge invariance.

Unfortunately, a $SU(2)_{L}$ gauge transformation leads to inconsistent results:
\begin{equation}
               H^g=U_L H, \quad H^{\dagger g} = H^\dagger U^{-1}_L+\frac{i}{g_2}\partial_{4}U^{-1}_L.
\end{equation}
Therefore, we cannot give the $x_4$ dependence to the gauge parameters. 
Because of the absence of the fifth components of the  $SU(3)$ and $U(1)$ gauge fields, 
these gauge parameters cannot have the $x_4$ dependence , either. 
Moreover, the model is not invariant under five-dimensional Lorentz transformations. 
These are clearly attributed to the left-right asymmetry.
So let us demand that 
$\displaystyle \int dx_{4} {\cal L}( x_{\mu},x_{4}) $ is invariant under four-dimensional 
Lorentz transformations and $SU(3) \times SU(2)_{L} \times U(1)$ gauge transformations 
of which the gauge parameters are not dependent on $x_4$. 
We will put this as a guiding principle to construct our model.

The interaction terms $ig_2 \bar{ \psi_{L} } \gamma^{4} H \psi_{R}+ig_2 \bar{ \psi_{R}} 
\gamma^{4} H^{\dagger} \psi_{L}$ are gauge invariant but the $\partial_{4}$ terms 
are not. Here, let us assume that the fifth dimension has curled up into a very 
small circle. A field defined in this periodic space must satisfy 
$\varphi(x_{\mu},x_{4})=\varphi(x_{\mu},x_{4}+2\pi R)$, 
which means that it can be power expanded in periodic eigenfunctions:
\begin{equation}
               \varphi(x_{\mu},x_{4})=\sum_{n} e^{ip x_{4}}\varphi_{n}(x_\mu) ,\quad p=\frac{n}{R},
\end{equation}
where $R$ is the radius of this space, and $n$ is an arbitrary integer.
We assume that all fields are constructed by only periodic eigenfunctions which 
satisfy
\begin{equation}
               \int^{2\pi R}_0 dx_{4}(\mbox{ the}\ \partial_{4}\ \mbox{terms}) =0.
\end{equation}
For example, $\psi_{L}(x_\mu,x_4)=e^{ipx_{4}} \psi_{L}(x_{\mu})$ and 
$\psi_{R}(x_\mu,x_4)=e^{-ipx_{4}} \psi_{R}(x_{\mu})$ satisfy
\begin{equation}
               \int^{2\pi R}_0 dx_{4}\bar{\psi_L}(x_\mu,x_4)\gamma^{4}\partial_{4}\psi_R(x_\mu,x_4) =0.
\end{equation}
In this sense, we can set
\begin{equation}
               \partial_{4} \rightarrow 0.
\end{equation}
Then Eq. (\ref{eqn:H}) becomes
\begin{equation}
               {\cal D}^{(\varphi)}_{4}=\left( 
               \begin{array}{cc}
                              0 & -i g H \\
                              -i g H^{\dagger} & 0
               \end{array}
               \right),
\label{eqn:b}
\end{equation}
and we have the fermionic part of the Lagrangian density, 
using Eqs. (\ref{eqn:Dmu}) and (\ref{eqn:b}),:
\begin{equation}
               {\cal L}_{f} = i \bar{\Psi}\gamma^{\alpha}{\cal D}_{\alpha} \Psi.
\end{equation}

Now let us consider the bosonic part of Lagrangian density. 
According to the procedure of Sogami\cite{rf:2}, we define field strengths 
${\cal F}^{(k)}_{\alpha\beta}\ (k=1,2,3)$ through the commutator of the covariant 
derivatives as follows:
\begin{equation}
               [{\cal D}^{(\varphi)}_{\alpha},{\cal D}^{(\varphi)}_{\beta}]
               =-i\sum^{3}_{k=1}\rho_{k}g_{k}{\cal F}^{(\varphi)(k)}_{\alpha\beta},
\end{equation}
where the $\rho_k$'s are necessary to normalize the kinetic terms of the gauge fields 
in the Lagrangian density. The field strengths are calculated as follows:
\begin{eqnarray}
               \rho_{1}{\cal F}^{(\varphi)(1)}_{\mu\nu}&=&
               \left(
               \begin{array}{cc}
                              \frac{1}{2}Y^{(\varphi)(1)}_{L}F^{(1)}_{\mu\nu} & 0 \\
                              0 & \frac{1}{2}Y^{(\varphi)(1)}_{R}F^{(1)}_{\mu\nu}
               \end{array}
               \right),
               \quad
               \rho_{1}{\cal F}^{(\varphi)(1)}_{\mu 5}=0, \nonumber \\
               \rho_{3}{\cal F}^{(q)(3)}_{\mu\nu}&=&
               \left(
               \begin{array}{cc}
                              \frac{ \mbox{\boldmath $\lambda$} }{2}\cdot
                              \mbox{\boldmath $F$}^{(3)}_{\mu\nu} & 0 \\
                              0 & \frac{\mbox{\boldmath $\lambda$} }{2}\cdot
                              \mbox{\boldmath $F$}^{(3)}_{\mu\nu}
               \end{array}
               \right),
               \quad
               \rho_{3}{\cal F}^{(q)(3)}_{\mu 4}=0, \nonumber \\
               \rho_{3}{\cal F}^{(l)(3)}_{\alpha\beta}&=& 0, \nonumber \\
               \rho_{2}{\cal F}^{(\varphi)(2)}_{\mu\nu}&=&
               \left(
               \begin{array}{cc}
                              \frac{ \mbox{\boldmath $\tau$} }{2}\cdot
                              \mbox{\boldmath $F$}^{(2)}_{\mu\nu} & 0 \\
                              0 & 0
               \end{array}
               \right), \nonumber \\
               \rho_{2}{\cal F}^{(\varphi)(2)}_{\mu 4}&=&
               \left(
               \begin{array}{cc}
                              0 & \frac{1}{2}
                              \left(D^{(\tilde{\phi})}_{\mu}\tilde{\phi},D^{(\phi)}_{\mu}\phi \right) \\
                              \frac{1}{2}\left(D^{(\tilde{\phi})}_{\mu}\tilde{\phi},
                              D^{(\phi)}_{\mu}\phi \right)^{\dagger} & 0
               \end{array}
               \right),
\end{eqnarray}
where
\begin{eqnarray}
               F^{(1)}_{\mu\nu}&=&
               \partial_{\mu}A^{(1)}_{\nu}-\partial_{\nu}A^{(1)}_{\mu}, \nonumber \\
               \frac{ \mbox{\boldmath $\tau$} }{2}\cdot \mbox{\boldmath $F$}^{(2)}_{\mu\nu}
               &=&
               \partial_{\mu}A^{(2)}_{\nu}-\partial_{\nu}A^{(2)}_{\mu}
               -ig_{2}[A^{(2)}_{\mu},A^{(2)}_{\nu} ], \nonumber \\
               \frac{ \mbox{\boldmath $\lambda$} }{2}\cdot \mbox{\boldmath $F$}^{(3)}_{\mu\nu}
               &=&
               \partial_{\mu}A^{(3)}_{\nu}-\partial_{\nu}A^{(3)}_{\mu}
               -ig_{3}[A^{(3)}_{\mu},A^{(3)}_{\nu} ],
\end{eqnarray}
and
\begin{eqnarray}
               D^{(\phi)}_{\mu} &=& \partial_{\mu}-ig_2 A^{(2)}_{\mu}
                              -ig_1\frac{1}{2} y_{\phi} A^{(1)}_{\mu}, \nonumber \\
               D^{(\tilde{\phi})}_{\mu} &=& \partial_{\mu}-ig_2A^{(2)}_\mu 
                              -ig_1\frac{1}{2} y_{\tilde{\phi} } A^{(1)}_{\mu}, \nonumber 
\end{eqnarray}
\begin{equation}
               y_{\phi}=1,\quad y_{\tilde{\phi}}=-1.
\end{equation}
Note that the hypercharges for $\phi$ and $\tilde{\phi}$ are properly fixed through these of 
fermion fields. We can derive the kinetic terms of the $SU(2)_L$ gauge fields and the Higgs 
field from the quadratic form of ${\cal F}^{(\varphi)(2)}_{\alpha\beta}$:
\begin{equation}
               \sum_{\varphi=q,l} 
               {\rm Tr}^{(\varphi)} 
               \left(
               {\cal F}^{(\varphi)(2)}_{\alpha\beta}{\cal F}^{(\varphi)(2)\alpha\beta}
               \right)
               =\frac{1}{\rho^{\ 2}_{2}}
               \left\{ 
               \frac{1}{4}\mbox{\boldmath$F$}^{(2)}_{\mu\nu} \cdot\mbox{\boldmath$F$}^{(2)\mu\nu}
               -\left(
               D^{(\phi)}_{\mu}\phi\right)^{\dagger} \left(D^{(\phi)\mu}\phi 
               \right) 
               \right\}.
\end{equation}
Note that 
$ {\cal F}^{(2)}_{\alpha\beta}{\cal F}^{(2)\alpha\beta}={\cal F}^{(2)}_{\mu\nu}
{\cal F}^{(2)\mu\nu}-2{\cal F}^{(2)}_{\mu 4}{\cal F}^{(2)\mu}_{\quad\  4}$
from the metric Eq. (\ref{eqn:a}). The trace is defined by
\begin{equation}
               {\rm Tr}^{(q)}(\cdots)\equiv 
               \frac{1}{2}{\rm tr}^{(LR)}
               [ \frac{1}{3}{\rm tr}^{(3)} 
               \{ \frac{1}{2}{\rm tr}^{(2)}(\cdots) \}],
               \quad
               {\rm Tr}^{(l)}(\cdots)\equiv 
               \frac{1}{2}{\rm tr}^{(LR)}\{ \frac{1}{2}{\rm tr}^{(2)}(\cdots) \},
\end{equation}
where ${\rm tr}^{(LR)}$ is the trace over the two-dimensional chiral space, 
${\rm tr}^{(3)} ({\rm tr}^{(2)})$ is the trace over the internal $SU(3) (SU(2)_L)$ 
symmetry matrices. We may add gauge and four-dimensional Lorentz invariant 
interaction terms of $\phi$ to the Lagrangian density. Finally, we have the bosonic 
part of Lagrangian density:
\begin{eqnarray}
               {\cal L}_{b} &=&
               -\sum_{\varphi =q,l} \sum^{3}_{k=1}
               {\rm Tr}^{(\varphi)} 
               \left(
               {\cal F}^{(\varphi)(k)}_{\alpha\beta}{\cal F}^{(\varphi)(k)\alpha\beta}
               \right)
               -V(\phi) \nonumber \\
               &=&
               -\frac{1}{4}\sum^{3}_{k=1} \mbox{\boldmath$F$}^{(k)}_{\mu\nu} 
               \cdot\mbox{\boldmath$F$}^{(k)\mu\nu}
               +\left(D^{(\phi)}_{\mu}\phi\right)^{\dagger} \left(D^{(\phi)\mu}\phi \right) 
               -V(\phi).
\end{eqnarray}

The masses of the gauge bosons and the fermions are generated by spontaneous 
symmetry breaking as the standard model. Arkani-Hamed and Schmaltz\cite{rf:3} 
suggest that the hierarchies present in the Yukawa couplings of the standard model 
can be explained as a result of the slight displacement of the standard model field 
wavefunctions inside a four-dimensional domain wall in a higher-dimensional space. 
The effective four-dimensional Yukawa coupling is a product of the fundamental 
higher-dimensional Yukawa coupling and the overlap of the field wavefunctions.
We will apply their mechanism to our model. We assume that $A^{(k)}_{\mu}\ (k=1,2,3)$ 
is $x_4$ independent which is consistent with $\partial_{4} \rightarrow 0$ and 
$\phi$ can be written as a product: $\phi(x_\mu,x_4)=\rho(x_4)\varphi(x_\mu)$, that is, 
\begin{eqnarray}
               A^{(k)}_{\mu}(x_\nu,x_4)&=&A^{(k)}_{\mu}(x_\nu),  \\
               \phi(x_\mu,x_4)&=&\sum_n \phi_n(x_\mu,x_4),
\label{eqn:phi}
\end{eqnarray}
where
\begin{equation}
               \phi_n(x_\mu,x_4)=a_n e^{i\frac{n}{R}x_4 }\varphi(x_\mu). 
\label{eqn:c}
\end{equation}
We consider a potential of the following form:
\begin{eqnarray}
               V&=&\sum_n \{ \mu^{2}_{n}\phi^{\dagger}_n \phi_n+\lambda(\phi^{\dagger}_n \phi_n)^{2} \}. 
\label{eqn:Vphi}
\end{eqnarray}
In the case $\mu^2_n<0$ and $\lambda>0$, the potential $V$ has its minimum at 
\begin{equation}
               \phi_n^{\dagger} \phi_n=-\frac{\mu^2_n}{2\lambda} \equiv |v_n|^{2}.
\label{eqn:d}
\end{equation}
Substituting Eq. (\ref{eqn:c}) into Eq. (\ref{eqn:d}), we obtain
\begin{equation}
               \varphi^\dagger(x_\mu) \varphi(x_\mu) =\left| \frac{v_n}{a_n} \right|^2 \equiv v_0^{\ 2},
\end{equation}
where $v_0$ is a constant independent of $n$.
We therefore can write
\begin{equation}
               \varphi(x_\mu)=\sqrt{\frac{1}{2} }
               \left(
               \begin{array}{c}
                              0\\
                              v_0+h(x_\mu) 
               \end{array}
               \right),
               \quad 
               v_0 a_n=i v_n,
\label{eqn:e}
\end{equation}
where $h(x_\mu)$ is a real field. Substituting Eq. (\ref{eqn:e}) into Eq. (\ref{eqn:phi}), 
we have
\begin{eqnarray}
               \phi (x_\mu,x_4)&=&
               \sum_n iv_n e^{i\frac{n}{R}x_4} 
               \sqrt{\frac{1}{2} }
               \left(
               \begin{array}{c}
                              0\\
                              1+\frac{1}{v_0}h(x_\mu) 
               \end{array}
               \right) \nonumber \\
               &=&
               \sqrt{\frac{1}{2} }
               \left(
               \begin{array}{c}
                              0\\
                              iv(x_4)(1+\frac{1}{v_0}h(x_\mu) )
               \end{array}
               \right),
\end{eqnarray}
where 
\begin{equation}
               v(x_4)\equiv \sum_n v_n e^{i\frac{n}{R}x_4}.
\end{equation}
$v_0$ is determined by the normalization of the kinetic term of $h(x_\mu)$ as 
\begin{equation}
               v_0^{\ 2}=\int^{2\pi R}_0 dx_4 v^{*}(x_4) v(x_4).
\end{equation}
Thus, the potential (\ref{eqn:Vphi}) derives a vacuum expectation value depending on $x_4$.

For the quarks, we take\cite{rf:4}
\begin{eqnarray}
               \Psi(x_\mu,x_4) &=& \sum_{j=1}^{3}
               \left(
               \begin{array}{c}
                              \zeta^j_L(x_4) \psi^j_L(x_\mu) \\
                              \zeta^j_R(x_4) \psi^j_R(x_\mu)
               \end{array}
               \right), 
\label{eqn:f}
\end{eqnarray}
\begin{equation}
               \int^{2\pi R}_0 dx_4 \zeta^{i*}_{h}(x_4) \zeta^{j}_{h}(x_4) =\delta_{ij},
               \quad
               h=R,L,
\end{equation}
where the superscripts $i$ and $j$ represent the generations of quarks and 
$\zeta^{i}_{h}$ do not have spin indices. Then, 
\begin{equation}
               \int^{2\pi R}_0 dx_4 i\bar{\Psi}\gamma^\mu {\cal D}^{(q)}_\mu \Psi
               =\sum^3_{j=1} \left( i\bar{\psi}^j_L \gamma^\mu D^{(q)L}_\mu \psi^j_L
               +i\bar{\psi}^j_R \gamma^\mu D^{(q)R}_\mu \psi^j_R \right),
\end{equation}
and
\begin{eqnarray}
               \lefteqn{ \int^{2\pi R}_0 dx_4 i\bar{\Psi}\gamma^4 {\cal D}^{(q)}_4 \Psi }
               \nonumber \\
               &=&
               \sum^3_{i,j} \int^{2\pi R}_0 dx_4 \frac{g_{2}}{2}
               \left\{ 
               \zeta^{i*}_L \bar{\psi}_L (i\tilde{\phi},i\phi) \zeta^{j}_R \psi_R 
               +\zeta^{i*}_R \bar{\psi}_R (i\tilde{\phi},i\phi)^{\dagger} \zeta^{j}_L\psi_L
               \right\} \nonumber \\
               &=&
               \sum_{i,j}^3 \left\{
               -\bar{u}^{i}_{L} M^{(u)}_{ij} u^{j}_{R} (1+\frac{h}{v_0})
               -\bar{d}^{i}_{L} M^{(d)}_{ij} d^{j}_{R} (1+\frac{h}{v_0})
               +{\rm h.c.}, \right\}
\end{eqnarray}
where
\begin{eqnarray}
               M^{(u)}_{ij}&=&-\frac{g_2}{2\sqrt{2} }
               \int^{2\pi R}_0 dx_4 v^{*}(x_4) \zeta^{i*}_{L}(x_4) \zeta^{j}_{R}(x_4), 
               \nonumber \\
               M^{(d)}_{ij}&=&\frac{g_2}{2\sqrt{2} }
               \int^{2\pi R}_0 dx_4 v(x_4) \zeta^{i*}_{L}(x_4) \zeta^{j}_{R}(x_4). 
\end{eqnarray}
Note that $\gamma_4 \psi_L=-i\psi_L$, $\gamma_4 \psi_R=i\psi_R$.
$M^{(u,d)}_{ij}$ can be interpreted as the usual mass matrices. 
By replacing $\Psi$ to the lepton field, the mass matrices of leptons are obtained. 
For a charged weak boson, we have
\begin{equation}
               M_W=\frac{1}{2}v_0 g_2.
\end{equation}
Note that it is always possible to make a relation $m_t/M_W >1$.

To summarize, the integration over the $x_4$ of the five-dimensional Lagrangian density
\begin{eqnarray}
               \int^{2\pi R}_0 dx_4 {\cal L}&=&\int^{2\pi R}_0 dx_4 
               \left[ 
               i \bar{\Psi}\gamma^{\alpha}{\cal D}_{\alpha} \Psi
               -\sum_{\varphi =q,l} \sum^{3}_{k=1}{\rm Tr}^{(\varphi)}  
               \left(
               {\cal F}^{(\varphi)(k)}_{\alpha\beta}{\cal F}^{(\varphi)(k)\alpha\beta}
               \right) \right.  \nonumber \\
               & &\left. -\sum_k \{ \mu^{2}_{k}\phi^{\dagger}_k \phi_k
                                             +\lambda(\phi^{\dagger}_k \phi_k)^{2} \} 
               \right]
\end{eqnarray}
reproduced the result of the standard model. Unfortunately, the gauge transformations 
of ${\cal L}$ cannot have the gauge parameters depending on $x_4$. Owing to this 
strong restriction, renormalization may not acquire the advantage of the extension to 
five dimensions. However, we consider it is worthwhile to point out that there is room 
which can introduce an $SU(2)_L$ doublet of scalar fields into the theory as a gauge field 
and we can explain why the Higgs field is an isospin doublet with weak hypercharge $y=1$. 
$SU(3) \times SU(2) \times U(1)$ invariant Lagrangian with a high degree of symmetry 
does not require the above restrictions at all. This may suggest that our five-dimensional 
standard model is the effective low energy theory of $SU(3) \times SU(2) \times U(1)$. 
It is debatable point.

%%%%%%%%%%%%%%%%%%%%%%%%%%%%%%%%%%%%%%%%%%%%%%%%%%%%%%%%%%%%%%%%%%%%%%%%%%%%%%

\end{document}